\documentstyle[aps,12pt,cite]{revtex}

\begin{document}

\title {The Instantonic Approach With Non-Equivalent Vacua \footnote{Presented
at the Seventh International Wigner Symposium. August 2001. College Park,
MD, USA.} } 

\author { J. Casahorr\'an }
\address{Departamento de F\'{\i}sica Te\'orica, \\
Universidad de Zaragoza, E-50009 Zaragoza, Spain \\
E-mail:javierc@posta.unizar.es }

\maketitle
\bigskip

\centerline {\bf Abstract}

\begin{abstract}

We study the quantum-mechanical tunneling phenomenon in models which
include the existence of non-equivalent vacua. For such a purpose we
evaluate the euclidean propagator between two minima of the potential
at issue in terms of the quadratic fluctuations over the corresponding
instantons. The effect of the multi-instanton configurations are included
by means of the alternate dilute-gas approximation.

\end{abstract}

\section {Introduction.}

The tunneling through classically forbidden regions represents one of the
most striking phenomenon in quantum theory and therefore plays a central
role in many areas of modern physics. On the other hand, together with the
operator formalism of quantum mechanics we have an equivalent description
by means of path-integrals. In such a case the Schrodinger's equation is
substituted by a global approach where the quantum mechanical time evolution
is analysed in terms of a functional integration. Qualitatively speaking, the
path-integral representation corresponds to a sum over all histories allowed
to the physical system we are dealing with. To be precise, we need to take
into account an imaginary exponential of the classical action associated
with every path which fulfill the appropiate initial and end points conditions.
Of course, the quantum amplitude so built is difficult to handle due to the
oscillating character of the exponential at issue. To avoid problems of this
sort we carry out the change $t \rightarrow -i \tau$ (known in the literature
as the Wick rotation). In doing so, we can take advantage of the euclidean 
version of the path-integral which represents by itself a new tool for describing
relevant aspects of the quantum theory. \par

Almost from the very beginning of the subject a semiclassical treatment of the
tunneling phenomena (ranging from periodic-potentials in quantum mechanics to
Yang-Mills models in field theories) has been performed by means of the
so-called instantons \cite{po}. Going to more physical terms, the instantons represent
localised finite-action solutions of the euclidean equation of motion. To be
specific, the euclidean equation is the same as the usual one for our particle
in real time  except that the potential is turned upside down. Although more
massive than the perturbative excitations, the instantons themselves become
stable since an infinite barrier separates them from the ordinary sector of
the model. The stability is reinforced by the existence of a topological
conserved charge which does not arise by Noether's theorem in terms of a well-behaved
symmetry, but characterizes the global behaviour of the system when the
imaginary time is large enough. Accordingly, it comes by no surprise that these
classical solutions have been considered in the literature under the name of
topological configurations. 
Once the appropiate classical solution is well-known, we make an expansion 
around the topological background to evaluate the quadratic fluctuations which
arise in terms of the functional determinant of a second order differential
operator. The integration is solved within the gaussian scheme except for
the zero-modes which appear due to the invariances of the model. To deal with 
these excitations we introduce collective coordinates so that ultimately the
gaussian integration is carried out only along the directions orthogonal
to the zero-modes. 
As a functional determinant includes an infinite product
of eigenvalues, the result should be in principle a highly divergent expression.
Fortunately we can regularize the fluctuation factors by means of the ratio
of determinants. \par

To start from scratch
let us describe  the instanton calculus for the one-dimensional
particle as can be found  in \cite{kl}. The  reader finds there
a comprehensive description of this subject. To be specific we assume
that our particle moves under the action of a confining
potential $V(x)$ which yields a pure discrete spectrum of energy eigenvalues.
If the particle is located at the initial time $t_i = -T/2$ at the point
$x_i$ while one finds it when $t_f = T/2$ at the point $x_f$, the
functional version of the non-relativistic quantum mechanics allows us to
express the transition amplitude in terms of a sum over all paths joining
the world points with coordinates $(-T/2, x_i)$ and $(T/2, x_f)$. 
Making the change $t \rightarrow - i \tau$, known in the
literature as the Wick rotation, the euclidean formulation of the
path-integral reads 

\begin{equation}
<x_f\vert \exp(- H T) \vert x_i> = N(T) \int [dx]  \ 
\exp\lbrace - S_e[x(\tau)]\rbrace
\label{eq:1}
\end{equation}

\noindent where $H$ represents the hamiltonian of the model, the
factor $N(T)$ serves to normalize the amplitude  while the symbol
$[dx]$ indicates the integration over all functions which fulfil the
boundary conditions at issue. Now we resort to the euclidean
action $S_e$, i.e.

\begin{equation}
S_e =  \int_{-T/2}^{T/2} \left[  {{1} \over {2}} 
\left( {{dx} \over {d\tau}} \right)^2 + V(x) \right]  \ d\tau
\label{eq:2}
\end{equation}

\noindent whenever the mass of the particle is set equal to unity for
notational simplicity. In this approximation we take for
granted the existence of a $x_c(\tau)$ configuration which represents
a stationary point of the euclidean action. Next we perform the expansion
of a generic $x(\tau)$ according to

\begin{equation}
x(\tau) = x_c(\tau) + \sum_j c_j \  x_j(\tau)
\label{eq:3}
\end{equation}

\noindent where as usual $x_j(\tau)$ stand for a complete set of orthonormal 
functions

\begin{equation}
\int_{-T/2}^{T/2} x_j(\tau) \  x_k(\tau) \ d\tau = \delta_{jk}
\label{eq:4}
\end{equation}

\noindent  vanishing at our boundary, i.e. $x_j(\pm T/2) = 0$. 
The eigenfunctions $x_j(\tau)$ (with eigenvalues $\epsilon_j$)
appear associated with the so-called stability equation given by

\begin{equation}
-  {{d^2 x_j(\tau)} \over {d\tau^2}}  +
 V^{\prime \prime}[x_c(\tau)] x_j(\tau) = \epsilon_j x_j(\tau)
\label{eq:5}
\end{equation}

To sum up, the euclidean transition amplitude reduces itself to

\begin{equation}
<x_f\vert \exp(- H T) \vert x_i> = N(T) \exp(- S_{eo}) \ \prod_j 
\epsilon_{j}^{-1/2}
\label{eq:6}
\end{equation}

\noindent where $S_{eo}$ represents
the classical action associated with the configuration $x_c(\tau)$ while the
product of eigenvalues is usually written as

\begin{equation}
\prod_j \epsilon_{j}^{-1/2} = \left\{ Det \left[
-  {{d^2} \over {d\tau^2}}  +
 V^{\prime \prime}[x_c(\tau)]  \right] \right\}^{-1/2}
\label{eq:7}
\end{equation}

\noindent according to the conventional notation of 
the finite-dimensional case. To fix the factor $N(T)$ we resort to a
well-known problem where

\begin{equation}
V(x) = {{\nu^2 } \over {2}} \  x^2
\label{eq:8}
\end{equation}

\noindent so that $V^{\prime \prime} (x = 0 ) = \nu^2$. As corresponds
to the harmonic oscillator the relevant amplitude is given by

\begin{equation}
<x_f = 0\vert \exp(- H_{ho} T) \vert x_i = 0> = N(T) \left\{ Det \left[
-  {{d^2} \over {d\tau^2}}  +
 \nu^2  \right]\right\}^{-1/2}
\label{eq:9}
\end{equation}

Next the evaluation of (\ref{eq:9}) is possible according to the general
method exposed in \cite{ra}. In other words

\begin{equation}
<x_f = 0\vert \exp(- H_{ho} T) \vert x_i = 0> = \left({{\nu} \over {\pi}}\right)^{1/2} \
\left(2 \sinh \nu T \right)^{-1/2}
\label{eq:10}
\end{equation}

\section {The sextic-model.}

As anticipated in the introduction the tunneling phenomenon appears as first
in the study of systems which have two or more degenerate classical minima
separated by potential energy barriers. The standard example would be the
point-particle in a double-well potential where, because of tunneling
precisely, the two first eigenstates are the spatially even and odd combinations
of harmonic oscillators centered at the bottom of the respective wells. In more
physical terms, the degeneracy of the energy eigenvalues is broken so that the
splitting term results proportional to the barrier-penetration factor. The
euclidean configuration which leads the process is of course the instanton whose
stability properties are decided by the behaviour of the small fluctuations
in its neighborhood. As a matter of fact the stability is assured by a clever
combination of translational invariance and conserved topological charges.
In the following we consider the triple-well potential $V(x)$ given by \cite{ca}

\begin{equation}
V(x) = {{\omega^2 } \over {8}} x^2 (x^2 - 1)^2
\label{eq:11}
\end{equation}

From a classical point of
view we find three minima located at $x_{-} = - 1$, $x_{+} =  1$ and
$x_{o} = 0$. Two of them, namely $x_{-}$ and $x_{+}$, are equivalent since
can be connected by means of the discrete symmetry $x \rightarrow - x$
at issue. However the third minimum $x_{o}$ is invariant under the action
of such a symmetry. In other words, the central vacuum is not identical
with the others located at both sides. 
When considering the limit  $\omega^2 \gg 1$ the energy
barriers are high enough to decompose the system into a sum of 
independent harmonic oscillators. However the existence of finite barriers
between the different wells of the potential yields a relevant
tunneling phenomenon so that ultimately the symmetry $x \rightarrow - x$ is
not spontaneously broken at quantum level and the expectation value of the
coordinate $x$ computed for the ground-state is zero as corresponds to the
even character of the potential $V(x)$. \par

\subsection {The one-instanton contribution.}

The question we wish to address at this point should be
 the explicit description of the tunneling in the euclidean version of the
path-integral.
As regards the one-instanton amplitude
we take into account the transition amplitude between 
$x_{o} = 0$ and $x_{+} =  1$. For such a purpose we need a classical configuration
with  $x_{i} = 0$ at $t_{i} = -T/2$ while
$x_{f} =  1$ when $t_{f} = T/2$. 
It is customarily assumed that $T \rightarrow \infty$ mainly because
the explicit solution of the problem is much more complicated for finite $T$. \par

First of all we can find the explicit form of the instanton $x_{c}(\tau)$
just by integration of a first-order differential equation, i.e.

\begin{equation}
 {{dx_{c}} \over {d\tau}} = \pm \sqrt{ 2 V(x_{c})}
\label{eq:12}
\end{equation}

\noindent where we recognize the quantum mechanical version of the Bogomol'nyi
condition \cite{bo}. 
Now we solve (\ref{eq:12}) by a simple quadrature so that

\begin{equation}
x_c(\tau) = \sqrt {[1 + \tanh \omega (\tau - \tau_c) ] / 2} 
\label{eq:13}
\end{equation}

\noindent where as usual the parameter $\tau_c$ indicates the point at which 
the instanton
makes the jump. As expected equivalent solutions are  obtained by means of the
transformation $\tau \rightarrow - \tau$ and $x_c(\tau) \rightarrow - x_c(\tau)$ so
that adjoint  minima of the potential can be connected by means of a topological
solution in a systematic way.
In addition we have that $S_{eo} = \omega/4$. On the other hand we
need classical configurations for which $x_{-} = 0$ and $x_{+} = 1$ at large
but finite values $\tau = \pm T/2$. However the explicit form of the instantons
that appear in the literature corresponds to infinite euclidean time. Fortunately
the difference is so small that can be ignored mainly because we are interested
in the limit $T \rightarrow \infty$. Our description of the
one-instanton amplitude between $x_{-} = 0$ and $x_{+} = 1$ takes over

\begin{eqnarray*}
<x_f = 1\vert \exp(- H T) \vert x_i = 0> = N(T) \left\{ Det \left[
- {{d^2} \over {d\tau^2}}  +
 \nu^2  \right]\right\}^{-1/2}   
\end{eqnarray*}
\begin{equation}
\left\{{{Det \left[- (d^2/d\tau^2) + V^{\prime \prime}[x_c(\tau)] \right]} \over
{Det \left[- (d^2/d\tau^2) + \nu^2 \right]}}\right\}^{- 1/2} \ 
\exp(-S_{eo})
\label{eq:14}
\end{equation}

\noindent where we have multiplied and divided by the determinant associated
with the harmonic oscillator of frequency $\nu$. As regards the determinant
built over the instanton itself we find a zero-mode $x_{o}(\tau)$ which could
jeopardize the computation procedure as a whole.
However this eigenvalue $\epsilon_o = 0$ comes by no surprise
since it reflects the translational invariance of the system. 
We can discover the
existence of this zero-mode starting from (\ref{eq:13}). Including the adequate
normalization one can check that

\begin{equation}
x_{o}(\tau) = {{1} \over {\sqrt{S_{eo}}}} {{dx_{c}} \over {d\tau}}
\label{eq:15}
\end{equation}

\noindent is just the solution of (\ref{eq:5}) with $\epsilon_o = 0$. 
The way out of this apparent cul-de-sac
is simple. The integration over $c_{o}$ (see (\ref{eq:3})) becomes equivalent
to the integration over the center of the instanton $\tau_{c}$. To fix the
jacobian of the transformation at issue we consider a first change like

\begin{equation}
\Delta x(\tau) = x_{o}(\tau) \  \Delta c_{o}
\label{eq:16}
\end{equation}

Going back to the general expression written in (\ref{eq:3}) we find that under
a shift $\Delta \tau_{c}$ the global effect corresponds to

\begin{equation}
\Delta x(\tau) = - \sqrt{S_{eo}} \ x_{o}(\tau) \Delta \tau_{c}
\label{eq:17}
\end{equation}

Finally the identification between (\ref{eq:16}) and (\ref{eq:17}) yields

\begin{equation}
d c_{o} = \sqrt{S_{eo}} \  d\tau_{c}
\label{eq:18}
\end{equation}

\noindent where the minus sign disappears since what matters is the modulus
of the jacobian at issue. To sum up we have that

\begin{eqnarray*}
\left\{{{Det \left[- (d^2/d\tau^2) + V^{\prime \prime}[x_c(\tau)] \right]} \over
{Det \left[- (d^2/d\tau^2) + \nu^2 \right]}}\right\}^{- 1/2} = 
 \end{eqnarray*}
\begin{equation}
\left\{{{Det^{\prime} \left[- (d^2/d\tau^2) + V^{\prime \prime}[x_c(\tau)] \right]} \over
{Det \left[- (d^2/d\tau^2) + \nu^2 \right]}}\right\}^{- 1/2} \
\sqrt{{{S_{eo}} \over {2 \pi}}} \ d\tau_{c}
\label{eq:19}
\end{equation}

\noindent where $Det^{\prime}$ stands for the well known reduced determinant
once the zero-mode has been removed. To make an explicit computation
of the quotient of determinants we resort to the  Gelfand-Yaglom
method where only the knowledge of the large-$\tau$ behaviour of the
classical solution $x_{c}(\tau)$ is necessary \cite{gy}. If $\hat{O}$ and
$\hat{P}$ represent a couple of second order differential operators, 
whose eigenfunctions
vanish at the boundary, the quotient of determinants is given in terms of the
respective zero-energy solutions $f_{o}(\tau)$ and $g_{o}(\tau)$ according to

\begin{equation}
{{Det \hat{O}} \over {Det \hat{P}}} = {{f_{o}(T/2)} \over {g_{o}(T/2)}}
\label{eq:20}
\end{equation}

\noindent whenever the eigenfunctions at issue satisfy the initial conditions

\begin{equation}
f_{o}(-T/2) = g_{o}(-T/2) = 0, \ \ \ {{df_{o}} \over {d\tau}} (-T/2) =
{{dg_{o}} \over {d\tau}} (-T/2) = 1
\label{eq:21}
\end{equation}

For the zero-mode $g_{o}(\tau)$ associated with the generic harmonic oscillator
of frequency $\nu$ we have

\begin{equation}
g_{o}(\tau) = {{1} \over {\nu}} \  \sinh [\nu (\tau + T/2)]
\label{eq:22}
\end{equation}

\noindent so that now we need the explicit form of the solution $f_{o}(\tau)$ which
corresponds to the topological configuration written in (\ref{eq:13}).
Starting from the $x_{o}(\tau)$ zero-mode we write a second solution
$y_{o}(\tau)$ given by 

\begin{equation}
y_{o}(\tau) = x_{o}(\tau) \ \int_{0}^{\tau} {{ds} \over {x_{o}^{2}(s)}}
\label{eq:23}
\end{equation}

In doing so we may summarize the asymptotic behaviour of $x_{o}(\tau)$ and
$y_{o}(\tau)$ as follows

\begin{equation}
x_{o}(\tau) \sim \left\{ \matrix{C \exp (- 2 \omega \tau) \ \ \ if \ \ 
\tau \rightarrow \infty \cr D \exp ( \omega \tau) \ \ \ \  if \ \ 
\tau \rightarrow - \infty \cr} \right.
\label{eq:24}
\end{equation}

\begin{equation}
y_{o}(\tau) \sim \left\{ \matrix{ \exp ( 2 \omega \tau)/4 \omega C \ \ \ if \ \
\tau \rightarrow \infty \cr - \exp (- \omega \tau)/2 \omega D \ \ \ \  if \ \
\tau \rightarrow - \infty \cr} \right.
\label{eq:25}
\end{equation}

\noindent where the constants $C$ and $D$ derive from the explicit form of 
the derivative of (\ref{eq:13}).  Starting now
from the linear combination of $x_{o}(\tau)$ and $y_{o}(\tau)$ given by

\begin{equation}
f_{o}(\tau) = A x_{o}(\tau) + B y_{o}(\tau)
\label{eq:26}
\end{equation}

\noindent the employ of the initial conditions  leads us to

\begin{equation}
f_{o}(\tau) = x_{o}(- T/2) y_{o}(\tau) - y_{o}(- T/2) x_{o}(\tau)
\label{eq:27}
\end{equation}

From this expression we can extract the asymptotic behaviour
of $f_{o}(\tau)$, i.e.

\begin{equation}
f_{o}(T/2) \sim {{D} \over {4 \omega C}} \ exp(\omega T/2) \ \ \ \  if  \ \ 
T \rightarrow \infty
\label{eq:28}
\end{equation}

Now we need to take into account the lowest eigenvalue of the stability
equation to obtain the value of the ratio of determinants. 
We can explain the situation as follows: the
derivative of the topological solution does not quite satisfy the boundary
conditions for the interval $(- T/2,T/2)$. When enforcing such a behaviour,
the eigenstate is compressed and the energy shifted slightly upwards. In such a
case the zero-mode $x_{o}(\tau)$ is substituted for the $f_{\lambda}(\tau)$ which
stands for 

\begin{equation}
- {{d^2 f_{\lambda}(\tau)} \over {d\tau^2}}  +
 V^{\prime \prime}[x_c(\tau)] f_{\lambda}(\tau) = \lambda f_{\lambda}(\tau)
\label{eq:29}
\end{equation}

\noindent whenever

\begin{equation}
f_{\lambda}(- T/2) = f_{\lambda}(T/2) = 0
\label{eq:30}
\end{equation}

Resorting now to the lowest order in perturbation theory we get

\begin{equation}
f_{\lambda}(\tau) \sim  f_{o}(\tau) + \left.\lambda \ {{df_{\lambda}} 
\over {d\lambda}} \right |_{\lambda = 0}
\label{eq:31}
\end{equation}

\noindent so that 

\begin{equation}
f_{\lambda}(\tau) =  f_{o}(\tau) + \lambda \int_{-T/2}^{\tau}
[x_{o}(\tau) y_{o}(s) - y_{o}(\tau) x_{o}(s)] \ f_{o}(s) \ ds
\label{eq:32}
\end{equation}

The asymptotic behaviour of $f^{o}(\tau)$, $x_{o}(\tau)$ and $y_{o}(\tau)$,
together with the condition $f_{\lambda}(T/2) = 0$ allows us to find this
lowest eigenvalue $\lambda$, i.e.

\begin{equation}
\lambda = 2 \omega D^2 \exp(- \omega T)
\label{eq:33}
\end{equation}

The Gelfand-Yaglom method provides us with the final expression
of the quotient of determinants whenever we choose for the frequency $\nu$ of
the harmonic oscillator of reference the average of the frequencies of the
central and lateral wells. In other words $\nu = 3 \omega /2$. We can compare
this situation with
 the well-grounded double-well model where the two minima of
the potential are equivalent so that the aforementioned average is out of order.
With this information we can write
the one-instanton amplitude between the points $x_{i} = 0$ and $x_{f} = 1$, i.e.  

\begin{eqnarray*}
<x_f = 1\vert \exp(- H T) \vert x_i = 0> = 
\end{eqnarray*}
\begin{equation}
 \left({{3 \omega} \over {2 \pi}}\right)^{1/2} \
\left(2 \sinh 3 \omega T/2 \right)^{-1/2} \  \sqrt{S_{eo}} \  
\sqrt{{{4} \over {3 \pi}}} \ 
\exp(-S_{eo}) \  \omega \  d\tau_{c} 
\label{eq:34}
\end{equation}

As expected we get a transition amplitude just
depending on the point $\tau_{c}$ at which the instanton makes
precisely the jump. Next we should take into account the configurations
constructed out of instantons and antiinstantons which mimic the behaviour
of a trajectory strictly derived from the euclidean equation of motion.
In doing so we get an additional bonus since the integration over the centers
of the string of instantons and antiinstantons is carried out in a systematic
way. But this point is the subject of the next subsection. \par

\subsection {The dilute-gas approximation.} 

Since the above calculations were carried out over a single instanton,
it remains to identify the  contributions associated with 
a string of widely separated 
instantons and anti-instantons along the $\tau$ axis.
It is customarily assumed that these combinations of topological solutions
represent no strong deviations of the trajectories just derived from the
euclidean equation of motion without any kind of approximation. We evaluate
the functional integral by summing over all such configurations, with
$k$ instantons and anti-instantons centered at points
$\tau_1,...,\tau_k$ whenever

\begin{equation}
-{{T} \over {2}} < \tau_1 < ... < \tau_k < {{T} \over {2}}
\label{eq:35}
\end{equation}

If the regions where the instantons (anti-instantons) make the
jump are narrow enough, the action of the
proposed path is almost extremal. In this approach
the total action  is
given by the sum of the $k$ individual actions. This scheme is well-known 
in the literature where it appears with the name of dilute-gas approximation
\cite{co}. 
The translational degrees of freedom of the  separated $k$
topological configurations yield an integral of the form

\begin{equation}
\int_{-T/2}^{T/2} \omega d\tau_k \
\int_{-T/2}^{\tau_k} \omega d\tau_{k - 1} ... 
\int_{-T/2}^{\tau_2} \omega d\tau_1 = {{(\omega T)^k} \over {k!}} 
\label{eq:36}
\end{equation}

As regards the quadratic fluctuations  we have
now that the single ratio of determinants transforms into 

\begin{eqnarray*}
\left({{3 \omega} \over {2 \pi}}\right)^{1/2} \
\left(2 \sinh 3 \omega T/2 \right)^{-1/2} \ 
\left\{{{Det^{\prime} \left[- (d^2/d\tau^2) + V^{\prime \prime}[x_c(\tau)] \right]} \over
{Det \left[- (d^2/d\tau^2) + 9 \omega^2/4 \right]}}\right\}^{- 1/2} \longrightarrow
\end{eqnarray*}
\begin{equation}
\left({{3 \omega} \over {2 \pi}}  \right)^{1/2} \exp(- 3 \omega T/4) \ 
\left[\left\{{{Det^{\prime} \left[- (d^2/d\tau^2) + V^{\prime \prime}[x_c(\tau)] \right]} \over
{Det \left[- (d^2/d\tau^2) + 9 \omega^2/4 \right]}}\right\}^{- 1/2}\right]^k 
\label{eq:37}
\end{equation}

\noindent according to the limit of the factor associated with the harmonic
oscillator when $T$ is large. 
When going to the dilute-gas approximation one considers a set of instantons
and anti-instantons so that each topological configuration starts where its
predecessor ends. For our amplitude  the total number $k$ of instantons
plus anti-instantons must be odd. The combinatorial
factor $F$ associated with the number of possible configurations corresponds to
$F = 2^{(k-1)/2}$ because the closed paths starting and coming back to the point
$x_{o} = 0$ require in a systematic way instanton-anti-instanton pairs.
We remind the difference with the double-well potential where the instantons
strictly alternate with the anti-instantons since the problem has only two
minima and the combinatorial factor is not necessary.
Now we can express the
complete transition amplitudes for the triple-well potential so that

\begin{equation}
<x_f = 1\vert \exp(- H T) \vert x_i = 0> = 
\left({{3 \omega} \over {4 \pi}}  \right)^{1/2} \exp(- 3 \omega T/4)
\sum_{j=0}^{\infty} {{(\omega T d)^{2j+1}} \over {(2j+1)!}}
\label{eq:38}
\end{equation}

\noindent where as usual $d$ stands for the  instanton density, i.e.

\begin{equation}
d = \sqrt{{{8} \over {3 \pi}}} \  \sqrt{S_{eo}} \  \exp(-S_{eo})
\label{eq:39}
\end{equation}

To sum up

\begin{equation}
<x_f = 1\vert \exp(- H T) \vert x_i = 0> = 
\left({{3 \omega} \over {4 \pi}}  \right)^{1/2} \exp(-3 \omega T/4) \
\sinh (\omega T d)
\label{eq:40}
\end{equation}

If we return to the euclidean transition amplitude written in (\ref{eq:1}),
the insertion of the pure discrete spectrum of energy eigenfunctions, namely

\begin{equation}
H \vert n > = E_n \vert n >
\label{eq:41}
\end{equation}

\noindent allows us to write that

\begin{equation}
<x_{f} = 1\vert \exp(- H T) \vert x_{i} = 0> = \sum_{n}
\exp(- E_n T) <x_{f} = 1 \vert n >
<n\vert x_{i} = 0 >
\label{eq:42}
\end{equation}

Next we denote by $E_{o}$, $E_{1}$ and $E_{2}$ the energies of the ground-state
and the two first excited levels of our problem. As the triple-well potential
we are dealing with is even we know that $<1\vert x_{i} = 0 >$ vanishes so that
the limit $T \rightarrow \infty$ in (\ref{eq:40}) provides us with the following
energy eigenvalues

\begin{equation}
E_{o} = {{3 \omega} \over {4}} - \omega d, \ \ \ \
E_{1} = {{3 \omega} \over {4}}, \ \ \ \ 
E_{2} = {{3 \omega} \over {4}} + \omega d
\label{eq:43}
\end{equation}

The tunneling  transfers along the whole real axis
the gaussian wave functions constructed over the points $x_{-} = - 1$,
$x_{+} =  1$ and $x_{o} = 0$ so that the splitting term results
proportional to the barrier-penetration factor. In addition the average of the harmonic
frequencies over the non-equivalent minima of the potential serves as the
central position for the splitting. \par


\begin{thebibliography}{99}

\centerline{\bf REFERENCES} \par

\bibitem{po}
{A. Polyakov,} Nucl. Phys. {\bf B120} (1977) 429.
\bibitem{kl}
{H. Kleinert,} {\it Paths Integrals in Quantum Mechanics, Statistics and
Polymer Physics}.
Singapore: World Scientific (1990).
\bibitem{ra}
{P. Ramond,} {\it Field Theory: A Modern Primer}.
New York: Addison-Wesley Publishing Company (1992).
\bibitem{ca}
{J. Casahorr\'an,} Phys. Lett. {\bf A283} (2001) 285.
\bibitem{bo}
{E. B. Bogomol'nyi,} Sov. Journ. Phys. {\bf 24} (1973) 1888.
\bibitem{gy}
{I.M. Gelfand and A.M. Yaglom,} Journ. Math. Phys. {\bf 1} (1960) 48.
\bibitem{co}
{S. Coleman,} {\it Uses of Instantons} in {\it The Whys of Subnuclear Physics}.
Ed. A. Zichichi.
New York: Plenum Press (1979).

\end{thebibliography}
\end{document}